\documentclass[sigconf]{acmart}
\AtBeginDocument{%
  }

\setcopyright{acmlicensed}
\copyrightyear{2026}
\acmYear{2026}
\acmDOI{10.1145/3822455.3838765}
\acmConference[MODELS 2026]{ACM / IEEE 29th International Conference on Model Driven Engineering Languages and Systems}{October 04--09,
  2026}{Malaga, Spain}
\acmISBN{978-1-4503-XXXX-X/2026/10}

\copyrightyear{2026}
\acmYear{2026}
\setcopyright{cc}
\setcctype{by-nc-nd}
\acmConference[MODELS 2026]{ACM/IEEE 29th International Conference on Model Driven Engineering Languages and Systems}{October 04--09, 2026}{Málaga, Spain}
\acmBooktitle{ACM/IEEE 29th International Conference on Model Driven Engineering Languages and Systems (MODELS 2026), October 04--09, 2026, Málaga, Spain}
\acmDOI{10.1145/3822455.3838765}
\acmISBN{979-8-4007-2809-9/2026/10}

\usepackage{graphicx}
\usepackage{svg}
\usepackage{amsmath}
\definecolor{green}{RGB}{76, 187, 23} 
\definecolor{yellow}{RGB}{255,165,0}     
\definecolor{red}{RGB}{255,0,0}

\AtBeginDocument{%

}

\begin{document}

\title{Towards an Asset Administration Shell Maturity Model}

\author{Carsten Ellwein}
\authornote{All authors contributed equally to this research.}
\authornote{Corresponding author.}
\email{{carsten.ellwein@isw.uni-stuttgart.de}}
\orcid{0000-0003-3286-4471}
\affiliation{%
  \institution{ISW, University of Stuttgart}
  \city{Stuttgart}
  \country{Germany}
}

\author{David Dietrich}
\authornotemark[1]
\email{{david.dietrich@isw.uni-stuttgart.de}}
\orcid{0000-0002-5558-2090}
\affiliation{%
  \institution{ISW, University of Stuttgart}
  \city{Stuttgart}
  \country{Germany}
}

\author{Rozana Cvitkovic}
\authornotemark[1]
\email{{rozana.cvitkovic@blueyonder.com}}
\orcid{0009-0001-8273-6278}
\affiliation{%
  \institution{Blue Yonder GmbH}
  \city{Stuttgart}
  \country{Germany}
}

\author{Andreas Wortmann}
\authornotemark[1]
\email{{andreas.wortmann@isw.uni-stuttgart.de}}
\orcid{0000-0003-3534-253X}
\affiliation{%
  \institution{ISW, University of Stuttgart}
  \city{Stuttgart}
  \country{Germany}
}

\renewcommand{\shortauthors}{C. Ellwein et al.}

\begin{abstract}
The Asset Administration Shell (AAS) is increasingly recognized as a fundamental model for the realization of and data exchange between digital twins in manufacturing. 
An AAS defines a hierarchical data structure to represent any type of asset throughout its entire lifecycle.
In the context of AAS-based systems, comparing different AAS instances constitutes a practical challenge, as neither a widely accepted methodological framework nor a maturity model are available to systematically support such analyses. 
To address this gap, we propose a novel concept of AAS maturity that characterizes the extent to which established digital twin criteria are met and thus enabling comparability of AAS instances. 
The concepts are derived from the literature and applied through exemplification. 
These emerging results enable practitioners and researchers to systematically compare AAS instances and support the identification and assessment of further development steps in the digital twin engineering process.
\end{abstract}




\maketitle


\section{Introduction}
\label{sec:Introduction}
Digital twins~\cite{kritzinger2018,tao2019five} are increasingly serving as the foundation for improved understanding, design, operation, and management of cyber-physical systems (CPS)~\cite{DJR+22}.
Although the software engineering community is largely responsible for developing the technologies that enable DTs~\cite{PLW+22}, manufacturing remains one of the main fields exploring their application~\cite{DJR+22}.
In manufacturing, a DT is a virtual counterpart of a real cyber-physical production system~(CPPS) \cite{EBC+21}, implemented and operated within a software environment~\cite{munoz2023conceptual}.
Different DTs perform distinct roles
, such as supporting analysis~\cite{827}, enabling control~\cite{1152}, or predicting behavior~\cite{splettstosser2023self}. 
To create DTs, various kinds of models have been proposed~\cite{lehner2025model,michael2025model}, ranging from UML~\cite{munoz2021using} and derivatives, over SysML~\cite{ferko2024towards} and AutomationML~\cite{lehner2021aml4dt}, to the Asset Administration Shell (AAS)~\cite{wagner2017role}.
%
The AAS is a key modeling technology to represent and implement DTs~\cite{Neubauer_2023} in manufacturing and has been successfully applied to, e.g., predictive maintenance~\cite{rahal2023asset}, model management~\cite{cavalieri2020asset}, and process control~\cite{dietrich2024}. 
It is conceived as the authoritative digitally instantiated representation of any type of asset throughout its entire lifecycle.  
Therefore, an AAS is specified as a hierarchical structure of data models, termed \textit{submodels}, each of which formalizes distinct components, properties, or functional aspects of the asset under consideration.
When AASs are developed collaboratively--such as through the integration of supplier-provided submodels into a comprehensive overall model or through the concurrent work of multiple engineering disciplines on a shared single source of truth--challenges emerge with respect to their mutual comparability, their maturity, and their applicability in terms of the intended use. 
A maturity model offers a systematic and effective framework for such comparisons~\cite{wendler2012maturity}.
However, no maturity frameworks for AAS are known~\cite{buttner2025}.
We aim to close this research gap.

Next, \autoref{sec:Background} illustrates the state-of-the-art on DTs and AASs, before \autoref{sec:DT} showcases the maturity of AASs, when perceived as DT in their runtime environment.
Afterward, \autoref{sec:Exemplification} illustrates the application of the maturity model. 
Based on this, \autoref{sec:Discussion} outlines limitations, \autoref{sec:RelatedWork} discusses related work, and \autoref{sec:FuturePlans} presents future plans. 
\autoref{sec:Conclusion} concludes.
\section{Background}
\label{sec:Background}


\subsection{Digital Twin}
\label{SdT:DT}

A common definition within academia distinguishes DTs from models and digital shadows based on their automated data flow \cite{kritzinger2018}. 
Although models are not connected to their physical counterparts, there is an automated connection between the physical object and its digital shadow or digital twin, for data exchange. 
For digital shadows, this exchange is unidirectional - the digital shadow automatically receives data from the physical object. 
In contrast, the digital twin closes the control loop and is capable of sending instructions back to its physical counterpart. 

Another academic definition focuses more on the structure of the DT than on the data flow between the digital object and its physical counterparts~\cite{tao2019five}.
According to this definition, referred to as the 5D model, a DT consists of: (i) the physical object itself, (ii) data from and about the physical object, (iii) models of the physical counterpart as well as models of the DT itself, (iv) services related to the physical object, and, as in the previous definition (c.f.~\cite{kritzinger2018}), the (v) connections between all elements of these dimensions.  
In this 5D model, every component is linked by bidirectional connections. 
Therefore, services can directly access real-time data from the physical system and issue commands to it. 
In addition, the services and the physical system can read, write, update, and delete the data and models provided by the DT.

A popular industrial definition of the DT is the Capabilities Periodic Table\footnote{\url{https://www.digitaltwinconsortium.org/initiatives/capabilities-periodic-table/}}~(CPT), published by the Digital Twin Consortium~(DTC), part of the Object Management Group~(OMG) and an association community within the US American non-profit trade association Enterprise Data Management~(EDM).
The CPT is characterized as a requirements definition framework that is independent of any specific architecture or technology.
According to the CPT, DTs can implement the following categories of capabilities: (i) \textit{Data Services}, (ii) \textit{Integration}, (iii) \textit{Intelligence}, (iv) user interaction (\textit{UX}), (v) \textit{Management}, and (vi) \textit{Trustworthiness}.  
Data services (i) include data management processes such as acquisition and ingestion, interpretation of data using ontologies, management of data through a model repository, and various data handling methods such as pub/sub, batch processing, and data aggregation.  
Integration (ii) refers to integration approaches that employ platforms, APIs, and enterprise systems to connect both higher-level, more abstract systems and lower-level, more detailed systems.  
Intelligence (iii) includes components and services for reasoning, planning, machine learning, simulation, and other capabilities that can be considered intelligent.
UX (iv) includes all system features that support monitoring, such as visualization of data, models, and their relationships, as well as interaction mechanisms including gamification and business intelligence methods, without directly affecting the machine.  
Management (v) functions are mainly concerned with the control and manipulation of the system. In this context, logging and data-related methods are advised for interpreting and configuring devices and machines. 
The trustworthiness (vi) capabilities comprise all features that contribute to the overall trustworthiness and reliability of the system.

Another, industrial driven definition is ISO 23247~\cite{ISO.23247}, an international standard that establishes a framework for DTs in manufacturing.
The standard identifies five elements of a DT framework: (i)~\textit{Data Services}, (ii)~\textit{System Components}, (iii)~\textit{Interfaces}, (iv)~\textit{Instances}, and (v)~\textit{Observable Manufacturing Elements}.
In contrast to the 5D model (cf.~\cite{tao2019five}), according to ISO 23247, the digital representation of the object (i.e., the DT) provides the interface for services to communicate with the physical counterpart. 

In summary, a DT in manufacturing can be described as a software system~\cite{munoz2023conceptual} that represents a  CPPS~\cite{kritzinger2018}, complements it with software services~\cite{tao2019five} and is connected, allowing bidirectional data exchange~\cite{kritzinger2018, tao2019five}.

\subsection{Asset Administration Shell}
\label{SdT:AAS}
In Industry 4.0 (I4.0), any element owned by an organization that contributes value to process execution is referred to as an asset~\cite{cavalieri2020asset}. 
Assets may be tangible, such as production equipment, workpieces, or even an entire factory, but can also be intangible, including models to describe machine behavior, software, or licenses~\cite{Frysak.2018}.

\begin{figure}[t]
\centering
    \includegraphics[width=0.75\columnwidth]{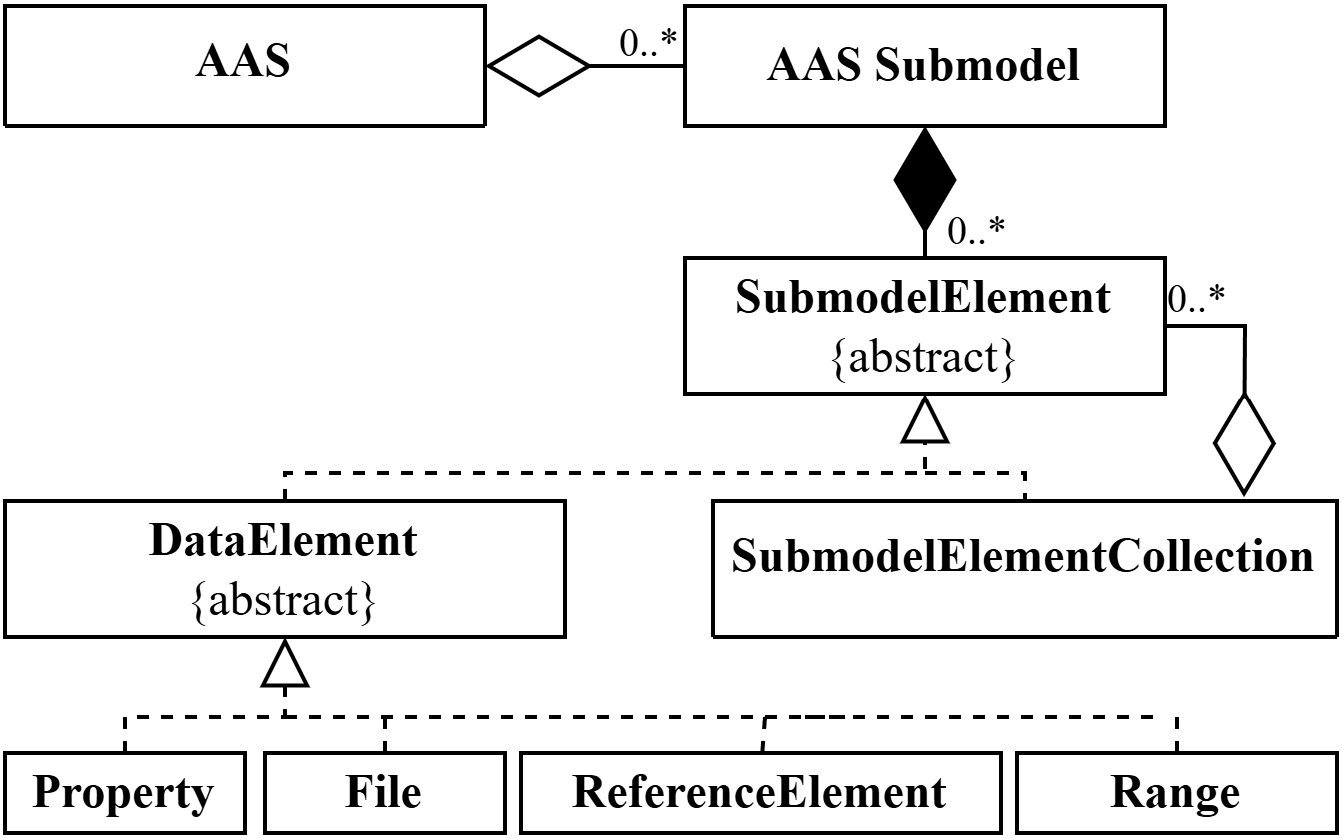}
    \caption{AAS metamodel~\cite{Zhang.2025}}
    \label{Fig:AASMeta}
\end{figure}


The AAS~\cite{wei2019review} is emerging as a potential standard for representing and implementing DTs~\cite{Neubauer_2023}, proposed as \textit{the definition and representation} layer within ISO 23247~\cite{ISO.23247}.
The development of AAS is driven by the Industrial Digital Twin Association (IDTA)~\cite{bader2019semantic}, an industry-oriented association related to the Association of German Mechanical Engineering Companies (VDMA). 
The AAS is used to model products~\cite{frick2024software}, processes~\cite{dietrich2024}, and resources~\cite{chen2025} and is intended to be the single source of truth digital representation of an asset throughout its entire life cycle~\cite{bader2022details}.

As the AAS contains relevant information throughout the lifecycle of an asset, it must be able to represent many types of content, including properties, modeled functions, parameters, an overview of integrated components, data generated during manufacturing or simulation, and descriptive information such as usage instructions and technical specifications. 
This requires the capability to store or reference heterogeneous data types~\cite{cavalieri2020asset}.

The structure of the AAS is specified by a metamodel (cf.~\cite{bader2022details}). 
The main subset of the class hierarchy related to the submodel structure of the AAS is illustrated in \autoref{Fig:AASMeta}.
At the highest level, the \textit{AssetAdministrationShell} class represents the entire AAS of an asset.
It can include multiple submodels, represented by the \textit{Submodel} class, each of which describes specific attributes of the asset. 
The individual attributes that make up a submodel — such as properties and files — are modeled by the abstract class \textit{SubmodelElement}.
The abstract class \textit{DataElement} is derived from \textit{SubmodelElement} and serves as a base for additional concrete classes. The classes \textit{Property}, \textit{Range}, \textit{File}, and \textit{ReferenceElement} model different kinds of asset attributes. 
The \textit{Property} encapsulates an information pair consisting of a single value and its data type, while \textit{Range} specifies two values together with their shared data type. 
\textit{File} denotes the type and location of a file, while \textit{ReferenceElement} defines a logical reference to either another element within the same or a different AAS, as well as external references.

To promote consistency and interoperability, the IDTA offers so-called submodel templates. These submodel templates are also standardized and made publicly accessible via their Content Hub\footnote{\url{https://industrialdigitaltwin.org/en/content-hub/submodels}}.
Every parameter within a submodel template is defined by an identifier, its semantics, and an illustrative example.  
The semantics are aligned with established dictionaries, such as the ECLASS\footnote{\url{https://eclass.eu/en/eclass-standard}} reference data standard for the unambiguous specification of products and services and the IEC Common Data Dictionary~(CDD), referred to as \textit{SemanticID}.  

\section{AAS Maturity}
\label{sec:DT}


In order to determine the maturity and thus the completeness of an AAS as a DT implementation, it is advisable to use the previously established definitions as criteria.
For this reason, the framework presented here is based on the well-known DT definitions data flow, 5D model, and the DTC CPT.
The five components of the 5D model~(cf.~\cite{tao2019five}), connections, data, services, physical entities, and virtual models serve as independent dimensions of maturity. 
A percentage value is calculated for each of these, representing the respective maturity.
All dimensions taken together then determine the overall academic maturity level of the AAS.

The maturity level $\mathcal{M}_C$ for \textit{connections} is based on data flows~\cite{kritzinger2018}, which describes a categorization of the connections to CP(P)S.
To derive connectable submodel elements, they are indicated by a custom concept qualifier of type ``Connection'' and the possible values ``Input'', ``Output'', or ``InOut''.
The connection of each element $i$ containing this qualifier type is then determined as
\begin{equation}
 c_i =
\begin{cases}
0 & \text{if no communication,} \\
0.5 & \text{if partial communication of InOut connections, and} \\
1 & \text{if the indicated communication}
\end{cases}
\end{equation}
is present.
The presence of such a connection can be derived from asset management systems or the submodel ``Asset Interfaces Mapping Configuration'' \cite{IDTA02027_2_0} (IDTA 02027).
The dimension of connectivity $\mathcal{M}_C$
for an AAS with $m$ connectable submodelelements $c_i$ is then calculated as the average.

For \textit{physical entities}, the maturity level $\mathcal{M}_E$ is distinguished between standalone assets and composed assets that contain subcomponents. 
For standalone assets the maturity is divided into three ordinal stages
and converted using the following linear scale:
\begin{equation}
   \mathcal{M}_{E_{single}} = \begin{cases}
        0\% & \text{if no physical asset exists at all,}\\
        50\% & \text{if a concept for the physical asset exists,}\\
        100\% & \text{if the physical asset is present.}
    \end{cases}
\end{equation}
For composed assets including a bill of material (BOM), the maturity level $\mathcal{M}_{E_{composed}}$
is calculated as the average based on its $n$ direct subcomponents $i$ indicated as $E_i$.
For multi stage BOMs this results in a recursive calculation.

The maturity level $\mathcal{M}_M$ of \textit{the virtual models} depends on how fully these models are completed.
To do so, the cardinality of the templates indicating required and optional values are used.
In the IDTA specification they are indicated by the qualifier ``SMT~/~Cardinality'' with its possible values ``ZeroToOne'', ``One'', ``ZeroToMany'' and ``OneToMany''.
Let $R$ be the set of required elements and $O$ the set of optional elements in submodels in an AAS.  
Let $r_i \in \{0,1\}$ indicate whether the required element $i$ is present 
and $o_j \in \{0,1\}$ whether the optional element $j$ is present.
For an existing AAS (or also submodels or submodel element collections), the maturity
\begin{equation}
    \mathcal{M}_{M}=B_M \left(w_{req} \cdot \frac{1}{|R|} \sum_{i \in R} r_i + (1-w_{req}) \cdot \frac{1}{|O|} \sum_{j \in O} o_j\right)
\end{equation}
is calculated based on the weighting $w_{req} \in [0,1]$ of required elements, as well as a further balance level $B_M$.
The weighting as well as the balance level need to be defined by users of the maturity framework based on the desired evaluation, e.g.  the weighting of optional values may be lower in early lifecycle phases.
As further submodels may be expected, but not yet added in the AAS, the balance $B_M = f(M_{SM,exp}, M_{SM})$, which compares existing and expected submodels as a percentage, is required to calculate the maturity level of the AAS.

The data within the models is examined to determine the maturity level $\mathcal{M}_D$ for the \textit{Data} component. 
The main focus here is on the actuality of the information, that is, whether the data can be updated.
Thus, the maturity
\begin{equation}
\mathcal{M}_D = \frac{M_{sme,con}}{M_{sme}}
\end{equation}

is calculated as a quotient of the number of all available submodel elements requiring a connection $M_{sme,con}$, as in the introduced qualifier of connectable elements, over the number of all available submodel elements $M_{sme}$ in the AAS.


\textit{Services} are diverse in nature, as they can, for example, provide data from the AAS, perform important functions for the runtime environment, or prepare information for end users.
This diversity creates challenges in determining the maturity level of services $\mathcal{M}_S$.
The DTC CPT shows the diversity of services and categorizes them into six different areas.

The category \textit{Data Services} (DS) includes services that deal with data processing and data management. 
Examples include data streaming, batch processing, and real-time processing. 
The data services have different approaches to data processing and are therefore mutually exclusive within a system, meaning that only one or a few services of this type would be present in the system at any given time, rather than all of them.
This leads to the assumption that the full maturity
\begin{equation}
    \mathcal{M}_{S_{DS}} = \max_{s_{DS,i} \in S_{DS}} s_{DS,i}
\end{equation}
is  reached as soon as one service $s_{DS,i} \in \{0,1\}$ exists without gradation in the set of Data Services $S_{DS}$ of DTC CPT.

The same applies to the next category, \textit{Integration} (IR). 
This category lists various services related to the integration of systems or platforms, such as API services, digital twin integration, and enterprise system integration. 
Since services are also used for different use cases here, fulfilling one criterion $s_{IR,i}$ in the set of integration services $S_{IR}$ is sufficient to achieve a maturity 
of 100\%, simular to the data services.

For the category \textit{Intelligence} (IC), the fulfillment of the first criterion plays a central role, whilst additional criteria may add further value. 
To reflect this, the maturity 
\begin{equation}
\mathcal{M_{S_{IC}}} = 1 - \frac{1}{2}^{k_{IC}}
\end{equation}
is described as a weighted score based on the number of fulfilled criteria $k_{IC}$ in the category.
Intelligence deals with data analysis and analytics, such as simulation, prediction, and reporting.

The same applies to the category \textit{UX}, which considers the presentation of information and topics that affect the end user, such as basic visualization, gamification, and dashboards.

For the category \textit{Trustworthiness} (TW), all criteria must be met for 100\% maturity.
A gradation is also possible if not all criteria are met. 
The maturity
\begin{equation}
\mathcal{M}_{S_{TW}} = \frac{M_{TW}}{N_{TW}}
\end{equation}
thus, is calculated based on the criteria fulfilled $M_{TW}$ out of the possible criteria $N_{TW}$ in the respective category.
Trustworthiness includes eight criteria related to system security, such as data encryption, security, safety, and privacy. 

This applies equally to the category \textit{Management} (MG), which deals with system monitoring. 
The category includes four services to achieve this in the best possible way, such as event logging and device management. 
Now, it has been explained how the individual maturity levels for the corresponding categories can be measured. 

To determine the overall maturity level of the services, the average of the six categories 
\begin{equation}
\mathcal{M}_S = \frac{1}{6} \sum_{i} \mathcal{M}_{S_i} \text{ for } i  \in \{DS,IR,IC,UX,MG,TW\}
\end{equation}
is calculated.
The maturity level for each of the components of the 5D model has now been assessed. 
To obtain the overall maturity level
\begin{equation}
\mathcal{M} = \frac{1}{5} \sum_{j} \mathcal{M}_{j} \text{ for } j \in \{C,D,E,M,S\}
\end{equation}
of the AAS, the average of the five sub-maturity levels must be calculated.

When displayed as a spider chart, the resulting maturity level is illustrated in \autoref{fig:ex:spiderweb}. 
In summary, the maturity level can provide an initial estimate of how close the AAS is to the most developed AAS in terms of DT requirements. 
This refers to an academic concept, recognizing that the most advanced AAS is not necessarily the most suitable choice for every particular use case.

\section{Exemplification}
\label{sec:Exemplification}

\begin{figure}[t]
    \centering
    \includegraphics[width=0.8\columnwidth]{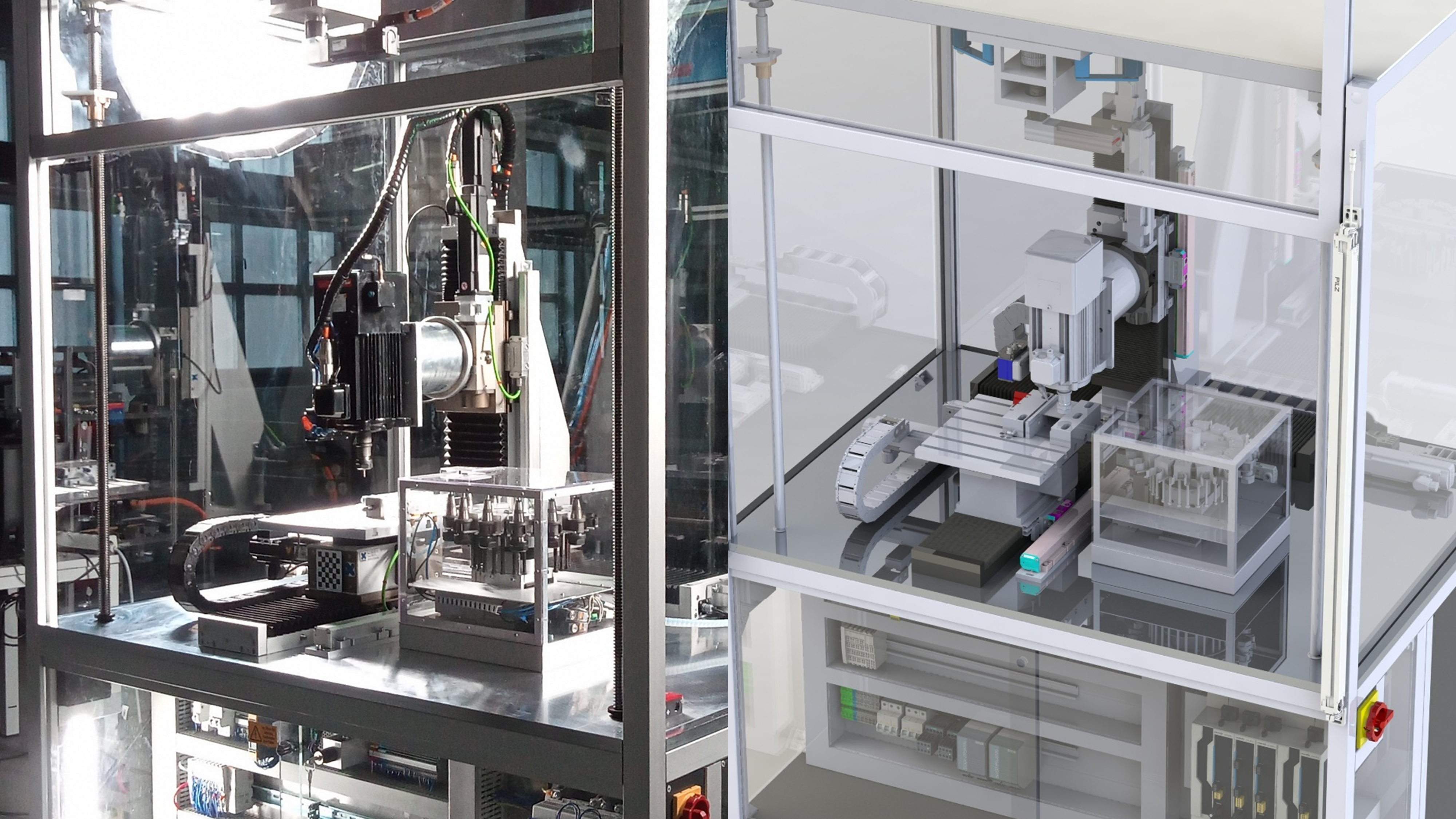}
    \caption{Five-axis milling machine at ISW (left) and its simulation model (right)}
    \label{fig:ex:fivex}
\end{figure}

To exemplify the proposed Maturity Model, we consider the engineering lifecycle of a five-axis milling machine controlled by Beckhoff TwinCAT (cf. \autoref{fig:ex:fivex}).
The scenario illustrates how maturity evolves across development phases and stakeholder perspectives prior to investment and operational deployment.

\begin{figure}[t]
    \centering
    \includegraphics[width=0.8\columnwidth]{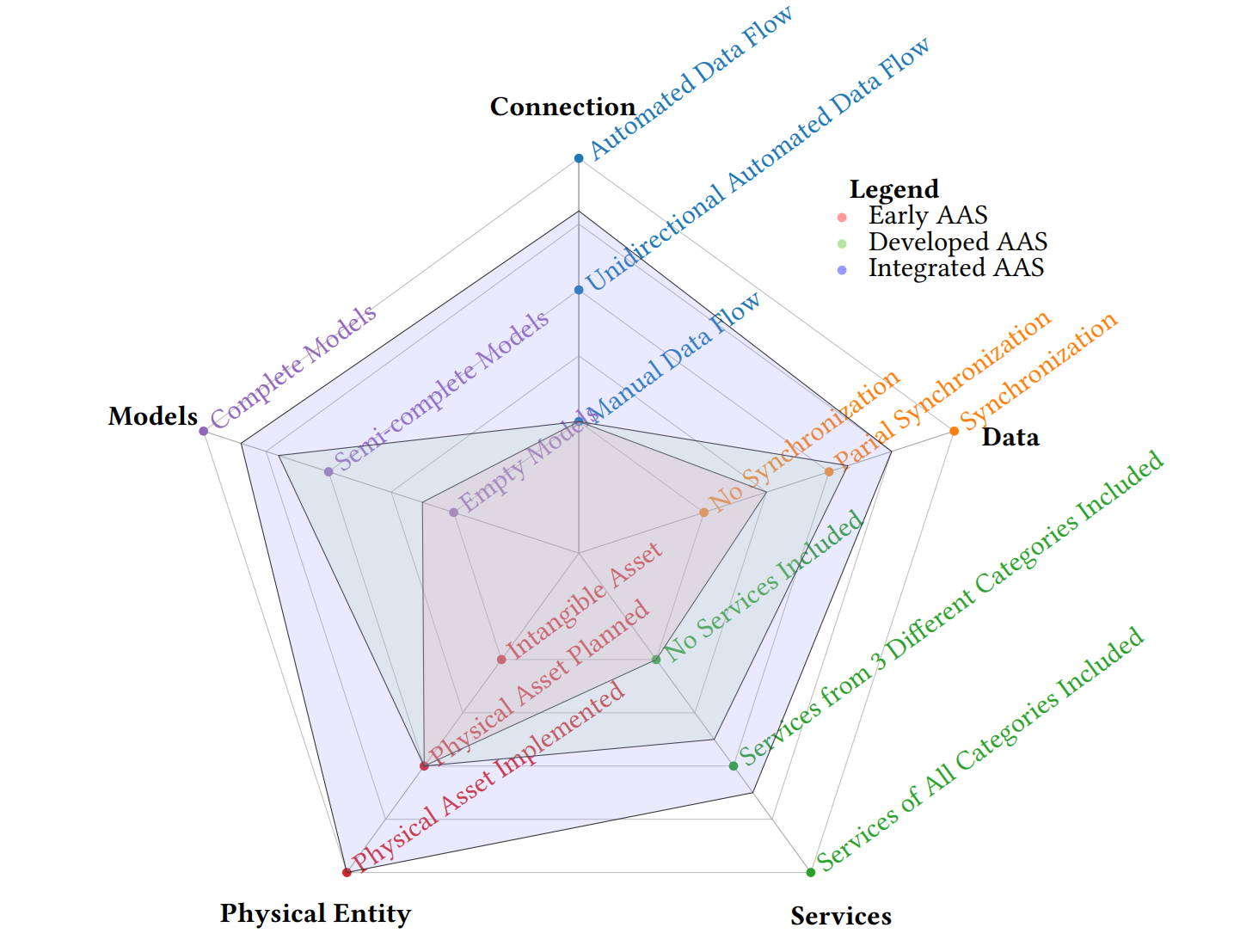}
    \caption{Maturity of the milling machine AAS throughout the engineering process}
    \label{fig:ex:spiderweb}
\end{figure}

During the early engineering phase, the manufacturer creates an initial standalone AAS.
It contains the standardized submodels ``Digital Nameplate'', containing 4 required and 54 optional elements,  and ``Technical Data'', containing 4 required and 21 optional elements at this development stage.
In order to address the defined application scenario of the AAS, 8 submodels are expected.
The ``Technical Data'' submodel documents planned feed rates and spindle speeds, as well as key construction parameters derived from preliminary calculations.
Although structurally compliant with the standardized submodel templates, the values are partly estimated and have not yet been validated against finalized drive configurations.
At that time, the submodel elements indicated as connections are the drive feed rates, spindle speed, and dimensions (9 optional elements), even if manually updated.
In contrast, 25 available elements (of which 8 are required) of the two templates do not contain a connection.
The maturity of the AAS at that time is shown in the inner red polygon in \autoref{fig:ex:spiderweb}:
As neither connection to the CPPS nor services exist, their corresponding maturity
$\mathcal{M}_C=\mathcal{M}_S=0$ equals zero.
In contrast, the data flow equals $\mathcal{M}_D=9/34\approx26.4\%$ due to the share of updated information.
Regarding the maturity of the models, the weighting $w_{req}=75\%$ is assumed to prioritize the required elements.
For the balance 
\begin{equation}
    B_M=\frac{1}{1+e^{\frac{M_{SM,exp}}{2}-M_{SM}}}
\end{equation}
a sigmoid function is used including the expected number of the submodels $M_{SM,exp}$ and the current number of submodels $M_{SM}$.
The sigmoid function is chosen to represent estimated maturity instead of a completion rate, where early progress is weighted less with an accelerating growth in the middle, when a significant number of submodels is present.
The maturity of models thus results in
$\mathcal{M}_M=\frac{1}{1+e^{0.5\cdot8-2}}(75\%*\frac{1}{4+4}*8+25\%*\frac{17+9}{54+21})=0.119*84\%\approx10\%$.
A high maturity can be determined, since all required data of the 2 submodels are present.
As the physical entity for the standalone AAS is planned, it results in maturity $\mathcal{M}_E=50\%$.
 The result is the overall maturity $\mathcal{M}=\frac{0+0+26.4+3.8+50}{5}\%\approx16\%$.

In the following engineering phase, the manufacturer completes the detailed design of the drives, leading to slightly modified technical parameters in the Technical Data submodel.
Furthermore, validated engineering artifacts are integrated in ``Handover Documentation'' (IDTA 02004), ``Provision of Simulation Models'' (IDTA 02005), the ``Provision of 3D Models'' (IDTA 02026), the ``Capability Description'' (IDTA 02020) and the ``Control Component Instance'' (IDTA 02016), resulting in an AAS ready to be delivered together with the resource instance.
The increased maturity of the AAS at this point is visible in the green polygon in \autoref{fig:ex:spiderweb}: 
As no physical asset exists, and thus automated data flow cannot exist, maturities in terms of connection and physical entity remain in mid-range.
In terms of services, an intelligent machine control (IR.IO, IC.CC) exists, the data have a basic visualization (UX.BV), and a system monitoring (MG.SM) is 
 available for the manufactured physical asset, resulting in service maturity $\mathcal{M}_S=\frac{0+1+0.5+0.5+0.25+0}{6}=37.5\%$. 
The data and model maturity are calculated similarly to the early AAS, resulting in $\mathcal{M}_D\approx57.5\%$ and $\mathcal{M}_M\approx70\%$


A customer requests the AAS of the milling machine for its integration in a production line.
Based on the evaluation, which has already been completed successfully, the customer can order the machine and plan its integration into the shop floor directly using the AAS maturity assessment.
Although structural- and simulation related submodels are present, operational integration requires bidirectional connectivity to its manufacturing execution system and an automated data synchronization of the AAS in its asset management system.
As the machine is delivered, the customer has already implemented these steps, resulting in the blue-shaded maturity shown in \autoref{fig:ex:spiderweb}  calculated analogously to the previous examples with a production line ready to use.
To reach full maturity in terms of services, the customer already plans a further data processing service and operation-accompanying digital twin for quality evaluation as well as improvements in terms of management and trustworthiness.
To achieve that, optional values will finally be integrated into its software systems leveraging the maturity of models and connections.

With this exemplification, the previously introduced maturity model is demonstrated.
The maturity model shows the increasing maturity over time as well as the necessary developments.
In combination, a structured evaluation for growing digital twins is shown.


\section{Discussion}
\label{sec:Discussion}

The introduced maturity model is based on simplifications in order to remain applicable in practice. 
Simplifying assumptions and the use of discrete scoring schemes may result in limited measurement resolution.  
It remains to be determined whether the AAS can, in practical applications, be differentiated with sufficient granularity.
Furthermore, the maturity model is based on the assumption that the AAS is intended to implement a DT. 
In other, theoretically possible use cases of the AAS, the validity of the maturity model is therefore very limited and may be misleading.
In particular, dependencies between dimensions can lead to incorrect assumptions in this context.
Based on the maturity model presented, statements regarding the applicability of AAS to a specific use case can only be made to a very limited extent. 

Although the maturity model was applied in the exemplification (cf.~\autoref{sec:Exemplification}) and several expert interviews were conducted during development, a systematic validation of the model has not yet been performed.
An initial theoretical differentiation from existing models is presented in the following subsection.

\section{Related Work}
\label{sec:RelatedWork}


Beyond its metamodel, the AAS specification~\cite{bader2022details} introduces three different types of AAS: \textit{(Type 1)} serialized as a file, \textit{(Type 2)} information retrieval through an API, and \textit{(Type 3)} peer-to-peer data exchange between AAS using interaction protocols specified in \cite{VDI2193-1.2020}. 
However, the specification does not provide additional details or criteria to differentiate between these types and their functionalities, which has led to different interpretations by different stakeholders~\cite{ellwein2026rethinking}. 
Whereas AAS type~1 is consistently characterized as a static or passive representation of asset-related information~\cite{13}, type~2 is described in various ways as an AAS that offers a common interface or a standardized API for data access~\cite{133}, as capable of reacting to external events~\cite{195}, as an AAS with decision-making capabilities~\cite{88}, or as an AAS that incorporates capabilities and skills~\cite{87}.
Type~3, in turn, is depicted as an AAS that interacts with other type~3 AAS~\cite{38}, enables communication using the I4.0 language~\cite{156}, includes service-oriented communication mechanisms~\cite{54}, or provides decision-making skills~\cite{13} with the ability to build multi-agent systems \cite{Sakurada.2023}.
In summary, types~2 and 3 exchange data with the physical object, thus extending the core functionality and coming close to the definition of a DT according to  data flows~\cite{kritzinger2018}, 5D model~\cite{tao2019five}, or ISO 23247~\cite{ISO.23247}. 
However, due to their poor differentiation, the two terms type~2 and type~3 can be used almost synonymously. 
AAS types therefore do not indicate anything about the maturity of the AAS.
The AAS portfolio analysis~\cite{ellwein2026rethinking} is based on the AAS types.
Although the resulting scheme yields an unambiguous categorization in contrast to the original AAS types, its applicability remains restricted to the specific mode of communication under consideration.
The application of the portfolio classification enables a preliminary assessment of maturity along the dimensions of "connection" and "data"; however, when considered in its entirety, it provides substantially less informative value than the model proposed in the present work.


The DTC recommends~\cite{DTC} applying Technology Readiness Levels (TRL)~\cite{heder2017nasa} to DTs and DT systems as a measure of maturity. 
The TRL provides information on the condition of the DT system~(e.g., the AAS), its stage of development, and industrial applicability. 
The suggestion to apply TRL is therefore certainly valid from the perspective of an industry association. 
However, TRL does not provide any information on the actual alignment of the AAS, and statements about stages of conceptual development cannot be inferred.
The TRL alone is therefore not sufficient to evaluate the overall maturity of the AAS.


Further models in the literature are the DTC CPT, as a capability model (cf.~\autoref{SdT:DT}) or the quantitative model~\cite{HU2023248}, which focuses on value, functions, and reliability, and evaluates performance based on 27 criteria. 
These models are broad, over-specified, and complex for practical applications.
The maturity model proposed in \cite{LIU2024102592} builds on the definition of DTs introduced in \cite{kritzinger2018} and extends it by introducing two additional levels: (i)~\textit{Cognitive DT} and (ii)~\textit{Federated DT}. 
The (i)~cognitive DT leverages algorithms, methods, and domain knowledge based on artificial intelligence~(AI) to autonomously generate and deliver feedback to the corresponding physical asset.
The (ii)~federated DT enables shared access to distributed data sources and connectivity interfaces, thus supporting the integration of advanced cross-domain DT applications.
Other DT maturity models follow similar approaches and conceptualize \textit{adaptive} and \textit{intelligent}~\cite{Madni.2019},
\textit{autonomous} and \textit{federated}~\cite{kim2020digital},
\textit{predictive}, \textit{optimized} and \textit{autonomous}~\cite{Masoumi02012023},
or \textit{self-adaptive} DTs~\cite{splettstosser2023self} as subsequent stages in the overall evolution of the DT paradigm.
The models, more suited to academic than practical applications, assume that the subject to be evaluated is indeed a DT. 
This assertion does not universally apply to all AAS; in certain instances, the AAS functions solely as a model~\cite{Sakurada.2023}. 
\section{Future Plans}
\label{sec:FuturePlans}

The subsequent section presents an overview of the roadmap for future research. 
This includes the (i) evaluation of the maturity model, (ii) weighting and tooling, as well as future work on the (iii) comparability and (iv) suitability of AAS instances. 

~\\\noindent\textbf{1. Empirical Evaluation:} %
As noted in the discussion of limitations~(cf.~\autoref{sec:Discussion}), this methodology is emerging research and has not yet been subjected to formal validation.
Future work will involve deploying the model across a range of application scenarios. 
To perform an evaluation, predictions are systematically assessed through expert interviews. 
Furthermore, the predictions will be benchmarked against established models (cf.~\autoref{sec:RelatedWork}).

~\\\noindent\textbf{2. Weighting and Tool Support:} %
The proposed maturity is calculated based on the weighting $w_{req}$ of required elements, as well as a further balance level $B_M$ (cf.~\autoref{sec:DT}). 
To enhance the comparability of the maturity levels achieved and reduce subjectivity, future research will focus on developing systematic design guidelines for weighting and balancing.
Furthermore, we intend to develop and integrate software-based tooling to systematically support parameter selection and the computation of maturity levels.

\pagebreak

\noindent\textbf{3. AAS Compatibility:} %
An issue that remains unresolved within the maturity model is the methodological comparison of two AAS instances with regard to the parameters they cover. 
Since AASs are complex constructs, it is necessary to define the level at which the comparison is to be made. 
Therefore, in future work, the hierarchical metamodel layout~\cite{flatscher2002} will be applied to the AAS.
On the basis of the findings, mathematical sum theory will be employed to systematically compare different AAS instances with respect to the proportional distributions observed at each metamodel layer.

~\\\noindent\textbf{4. AAS Suitability:} %
Another aspect, beyond the focus of the maturity model, is the concrete applicability or suitability of an AAS instance to serve as a data basis for a dedicated application. 
The suitability check will be targeted in future work. 
The approach will be a reference-based suitability assessment, where the application developer must provide a reference or template AAS that represents the minimum requirement profile.
The assessment will be carried out along four central dimensions that address different aspects of the conformity between a given AAS and the requirements of a specific application: (i)~structural conformity, (ii)~semantic consistency, (iii)~cardinality, and (iv)~conformity to specification. 


\section{Conclusion}
\label{sec:Conclusion}

This paper presents a method for assessing the maturity of the AAS in its runtime environment.
The applicability of the method is substantiated through illustrative examples.
The necessity of the model introduced is demonstrated in comparison with the related work.
The proposed maturity model provides a formalized and quantifiable framework. 
It provides a general overview of the extent to which the AAS actually functions as a DT. 
This will impact the further acceptance and interchangeability of AAS technology, as it can provide insights for further development and is the first step towards a structural comparison of the implementation status of two AASs.
Furthermore, the future plans present a roadmap for continued research, encompassing both AAS comparability and AAS suitability.
\balance

\begin{acks}
Partly funded by the Federal Ministry for Economic Affairs and Energy (BMWE) through the projects growING (grant no. 13IPC036G).
Partly funded by the German Federal Ministry of Research, Technology and Space (BMFTR) within the ``Research Campus – Public-Private Partnership for Innovation'' funding initiative (02P23Q820) and managed by the Project Management Agency Karlsruhe (PTKA).
\end{acks}

\bibliographystyle{ACM-Reference-Format.bst}
\bibliography{references}

\end{document}